\documentclass[prb,aps,twocolumn,floats,showpacs,superscriptaddress]{revtex4}
\usepackage{graphicx}

\newcommand{\be}{\begin{equation}}
\newcommand{\ee}{\end{equation}}
\newcommand{\bea}{\begin{eqnarray}}
\newcommand{\eea}{\end{eqnarray}}

\begin{document}
\title{Non-activated transport of strongly interacting two-dimensional holes in GaAs}
\author{Jian Huang}
\affiliation{%
Department of Electrical Engineering,
Princeton University, Princeton, New Jersey 08544, USA\\}%
\author{D. S. Novikov}
\author{D. C. Tsui}
\affiliation{%
Department of Electrical Engineering,
Princeton University, Princeton, New Jersey 08544, USA\\}%
\author{L. N. Pfeiffer}%
\author{K. W. West}%
\affiliation{
Bell Laboratories, Lucent Technologies, Murray Hill, New Jersey 07974, USA}
\begin{abstract}
We report on the transport measurements of two-dimensional holes in
GaAs field effect transistors with record low densities down to
$7\times10^{8}$ cm$^{-2}$. Remarkably, such a dilute system (with
Fermi wavelength approaching $1\,\mu$m) exhibits a non-activated
conductivity that grows with temperature approximately as a power
law at sufficiently low temperatures. We contrast it with the
activated transport found in more disordered samples and discuss
possible transport mechanisms in this strongly-interacting regime.

\end{abstract}

\maketitle
The question of how a strong Coulomb interaction can qualitatively
alter an electronic system is fundamentally important. It has
generated great interest in studying the transport in
two-dimensional (2D) electron systems.\cite{AFS} Since the
interaction becomes effectively stronger with lower 2D electron
density, samples with most dilute carriers are desirable to probe
the interaction effects. As the density is lowered, strong enough
disorder can localize the carriers so that the interaction effect is
smeared by the insulating behavior. Therefore, a clean 2D
environment is vital to uncover the underlying interaction
phenomena.

For a long time, the dilute 2D carriers have been known as
insulators characterized by activated conductivity. Specifically, in
an Anderson insulator,\cite{Anderson'58,stl} the conductivity
follows the Arrhenius temperature dependence $\sigma \sim
e^{-E_g/k_{B}T}$, where $E_g$ is the mobility edge with respect to
the Fermi level. The energy relaxation due to phonons in the
impurity band results in a softer exponential dependence $\sigma\sim
e^{-(T^*/T)^\nu}$, realized via the variable-range hopping (VRH)
process.\cite{Mott-VRH,ES} Here, the exponent $\nu=1/3$ for
non-interacting electrons,\cite{Mott-VRH} while $\nu=1/2$ if the
Coulomb gap opens up at the Fermi level.\cite{ES} Finally, strong
Coulomb interactions are believed to crystallize the 2D system
\cite{wc} which then can become pinned by arbitrarily small
disorder. A relation $d\sigma/dT>0$, being a natural consequence of
the activated transport, eventually became a colloquial criterion of
distinguishing an insulator from a metal.\cite{AKS-review}

The experimental results in the dilute carrier regime are known to
be greatly influenced by the sample quality, which has much improved
over time. The phonon-assisted hopping transport was observed in
early experiments.\cite{AKS-review} As the sample quality improved,
the later experiments performed on 2D electrons in cleaner
Si-MOSFETs demonstrated that the temperature-dependence of the
resistivity $\rho=\sigma^{-1}$ can be either metal-like
($d\rho/dT>0$), or insulator-like ($d\rho/dT<0$), depending on
whether the carrier density $n$ is above or below a critical value
$n_c$.\cite{mit} On the insulating side, where $n<n_{c}$, $\rho(T)$
grows exponentially with cooling.\cite{Mason} Similar results have
since been observed in various low disorder 2D systems, and the
resistivity on the insulating side has been consistently found to
follow an activated pattern $\rho\sim e^{(T^*/T)^\nu}$, with $\nu$
varying between $1/3$ and $1$.

In this work, we focus on the transport properties of clean 2D holes
in the dilute carrier regime where the insulating behavior is
anticipated. To achieve high quality and low density, we adopt the
GaAs/AlGaAs heterojunction insulated-gate field-effect transistor
(HIGFET) where the carriers are only capacitively induced by a metal
gate.\cite{kane,lilly,noh} Because there is no intentional doping,
the amount of disorder is likely to be less, and the nature of the
disorder is different from that of the modulation-doped samples.
Previous experiments on similar 2D hole\cite{noh,noh1} HIGFET
devices have demonstrated a non-activated transport. The temperature
dependence $\sigma(T)$ of the conductivity becomes approximately
linear, $\sigma \propto T$, \cite{noh1} when the density is lowered
to a minimum value of $1.6\times10^{9}$\,cm$^{-2}$. However, it is
unclear whether the linear $T$-dependence will persist for lower
densities or it is a crossover to a different transport regime.

We have measured several high quality $p$-channel HIGFET samples.
The hole density $p$ in our devices can be continuously tuned to as
low as $p=7\times10^{8}$ cm$^{-2}$, in which case the nominal Fermi
wavelength $\lambda_F = (2\pi/p)^{1/2} \simeq 0.95\,\mu$m. Our main
finding is that the conductivity $\sigma(T)$ of the cleanest samples
decreases with cooling in a non-activated fashion for densities down
to $7\times10^{8}$ cm$^{-2}$. The temperature dependence of the
conductivity appears to be best approximated by a non-universal
power-law $\sigma \propto T^{\alpha}$ with
$1\lesssim\alpha\lesssim2$. The systematic analysis of this
dependence will be published elsewhere.\cite{jian} At base
temperature, the magnitude of $\sigma$ is much greater than that of
a typical insulator with similar carrier density. Our results point
at the presence of the delocalized states in the system with a
record low carrier density. Thus our system is not an insulator even
though $d\sigma/dT>0$.

\begin{figure}[t]
\includegraphics[totalheight=3.5in, trim=0.15in 0.05in 0.15in 0.in]{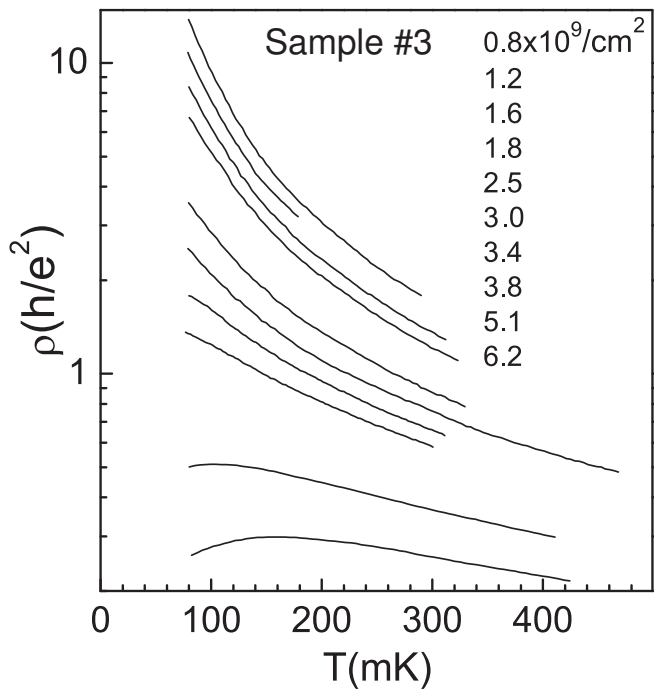}
\caption{\label{fig:rt} Temperature dependence of the resistivity of
sample \#3 in semi-log scale for a set of specified hole densities
listed on the righthand side.}
\end{figure}

\begin{figure}[t]
\includegraphics[totalheight=3.5in, trim=0.15in 0.005in 0.05in 0.in] {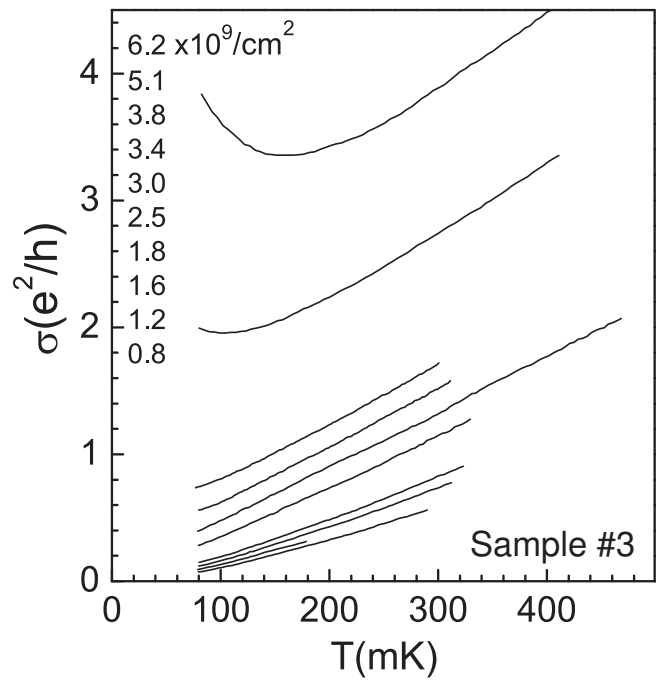}
\caption{\label{fig:ct} Re-plot of the data from Fig.~\ref{fig:rt}
in terms of conductivity vs. temperature.}
\end{figure}


The device geometry is a standard 3\,mm$\times 0.8$\,mm Hall bar.
The measurements were performed in a dilution refrigerator with a
base temperature of 35 mK. At each value of the gate voltage, the
mobility and density were determined through measuring the
longitudinal resistivity $\rho$ and its quantum oscillations in the
magnetic field. The temperature dependence of the resistivity was
measured with an ac four-terminal setup at high carrier density,
while both ac and dc setups were used for the low-density
high-impedance cases. To ensure linear response, current drive as
small as 1\,pA was used during the measurements at the lowest
carrier density. Driving currents of different amplitudes were used
for the low density cases and the measured resistivity did not
change with varied current drive.

The temperature dependence of the resistivity $\rho(T)$ for a number
of hole densities from sample \#3 is shown in Fig.~\ref{fig:rt},
with lowest temperature of $80$ mK. At first glance, the density
dependence of the $\rho(T)$ curves is similar to that found around
the metal-to-insulator transition, \cite{AKS-review} with
$p_c=4\times10^{9}$\,cm$^{-2}$ being the critical density. For
$p>p_c$, the system exhibits the apparent metallic behavior
($d\rho/dT>0$) at sufficiently low temperatures. The downward
bending of $\rho(T)$ becomes weaker as $p$ approaches $p_c$, and
disappears for lower $p$. The derivative $d\rho/dT$ at low $T$ then
becomes negative, a conventional characteristic of an insulator,
\cite{AKS-review} for the whole temperature range. At the
transition, the resistivity is of the order of $h/e^{2}$. The value
of $p_{c}$ is very close to that obtained in a similar device in
Ref.~\onlinecite{noh}.



Fig.~\ref{fig:ct} shows the conductivities $\sigma(T)$ of the same
sample (\#3) for the corresponding densities. For
$1.8\times10^{9}$\,cm$^{-2}<p<3.8\times10^{9}$\,cm$^{-2}$, the
conductivity increases approximately linearly with $T$ at high
temperatures (above $\sim 200$\,mK). The linear regions are almost
parallel for different densities, similar to that observed
previously.\cite{noh,noh1} This linear dependence occurs at
temperatures above the nominal Fermi temperature and will be studied
in detail elsewhere.\cite{jian} However, for lower densities, from
$8\times10^{8}$\,cm$^{-2}$ to $1.8\times10^{9}$\,cm$^{-2}$, the
conductivity deviates from the linear decrease at low temperatures.
$\sigma(T)$ exhibits a slower change with $T$ as the density is
reduced, with weaker $T$-dependence close to the base temperature.
The conductivity values ($\sim 0.1e^{2}/h$) are considerably larger
than those found in the more disordered sample which will be
described later.
\begin{figure}[t]
\includegraphics[width=3.4in]{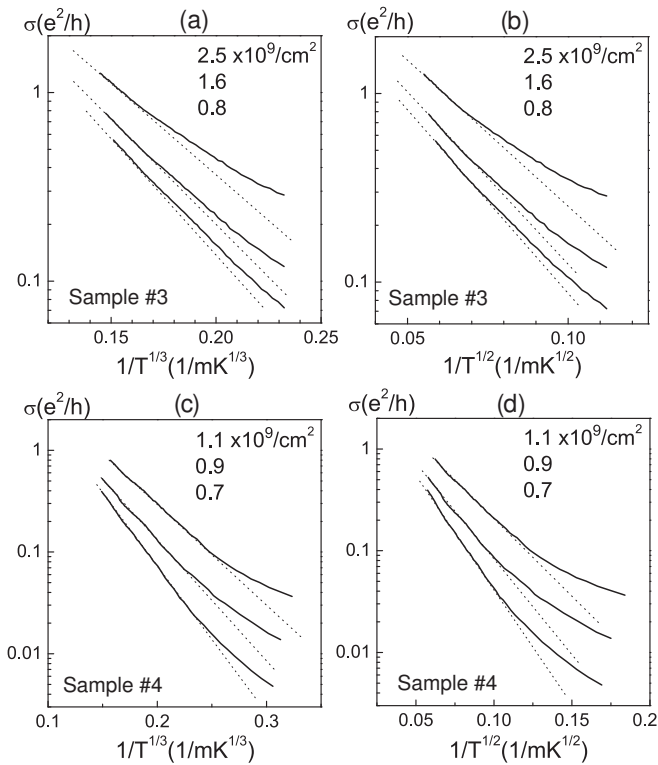}
\caption{\label{fig:VRH} Comparison of the conductivities
$\sigma(T)$ with VRH transport models: (a), (c) Mott; (b), (d)
Efros-Shklovskii. Panels (a) and (b) are results from sample \#3,
and (c) and (d) are from sample \#4.}
\end{figure}

In Fig.~\ref{fig:VRH}, we compare the measured conductivities with
the VRH predictions according to Mott \cite{Mott-VRH} and to Efros
and Shklovskii \cite{ES} for sample \#3 [panels (a) and (b)], and
sample \#4 [panels (c) and (d)]. The hopping conductivity
$\sigma\sim e^{-(T^*/T)^\nu}$, repeatedly observed in previous
experiments on the insulators, is expected to occur at low
temperatures. However, in both of our samples, the conductivity is
approximately linear (in the semi-log scale) at high temperatures
but nonlinear at low temperatures. It clearly deviates from the VRH
law (dotted lines), for both $\nu=1/3$ [panels (a) and (c)] and
$\nu=1/2$ [panels (b) and (d)]. The increasing deviation with
cooling indicates that the temperature dependence is weaker than
activated. The deviation is slightly larger for the E-S ($\nu=1/2$)
case.


\begin{figure}[t]
\includegraphics[width=3.4in, trim=0.07in 0.05in 0.02in 0in]{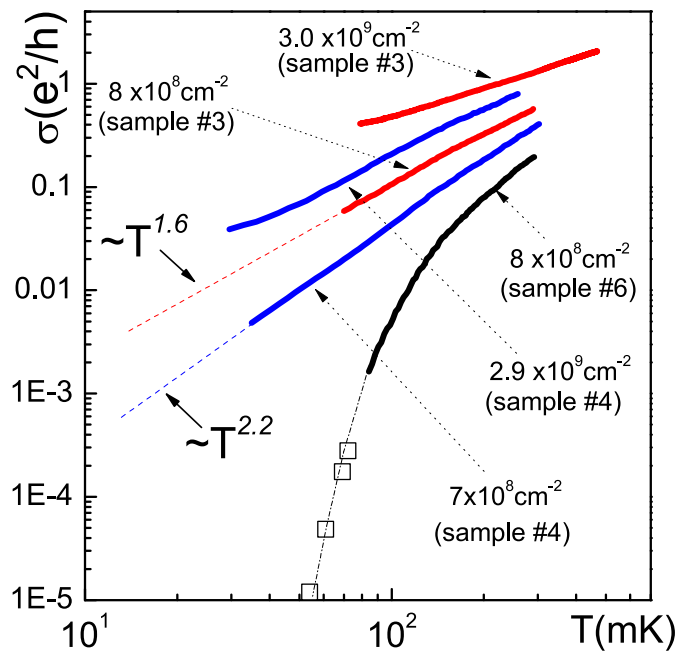}
\caption{\label{fig:log} Power-law $\sigma(T)$ behavior from clean
samples, samples \#3 and \#4, in comparison with the activated
behavior found in a more disordered sample (\#6). The scattered
points in the graph are the DC results of sample \#6.}
\end{figure}

The qualitative difference in the temperature dependence between our
clean samples from more disordered samples (previously measured) is
apparent in the log-log scale plot in Fig.~\ref{fig:log}. Here, the
conductivity from a more disordered sample (\#6) is also included
for comparison. For about the same densities ($7-8 \times
10^8\,$cm$^{-2}$), the dependence $\log\sigma$ versus $\log T$ for
the two cleanest samples appears to be approximately linear below
150\,mK, indicating a power-law-like relationship $\sigma \propto
T^{\alpha}$. The exponent $\alpha$, which corresponds to the slope
in the plot, is $\alpha \simeq 2.2$ for $p=7 \times 10^8$\,cm$^{-2}$
and $\alpha \simeq 1.6$ for $p=8 \times 10^8$\,cm$^{-2}$. Both
numbers differ from the much lower values previously observed in
carrier densities around $1.6\times 10^9$\,cm$^{-2}$.
\cite{lilly,noh1} The trend, larger $\alpha$ for lower density, is
consistent with the results in Ref.~\onlinecite{lilly}. On the other
hand, the conductivity for the more disordered sample (\#6) exhibits
a clear downward diving, consistent with the activated behavior.
Note that the low-$T$ conductivity in the more disordered sample is
at least three orders of magnitude smaller than that in the clean
ones for the same carrier density.

The insulating character of our more disordered sample is consistent
with previously observed Anderson insulators in lower quality 2D
systems. However, our clean HIGFET results show that, with less
disorder, the transport becomes non-activated, indicating the
presence of delocalized states.

What is responsible for the apparent delocalization?
The Anderson localization, being an interference effect,
can occur only when the system is sufficiently phase-coherent.
At finite temperature, strong electron-electron interaction
can dramatically reduce the
coherence length $l_\phi$. As the carrier temperatures in our case
are of the order of the Fermi temperature $T_F$, there is no suppression of the
interaction between the quasiparticles associated with the filled Fermi sea.
In this situation it is plausible to assume that $l_\phi \sim \lambda_F$.\cite{NZA}
On the other hand, the 2D localization length $\xi$ is
exponentially sensitive to the amount of disorder.\cite{loc-length}
Thus it seems likely for the phase coherence
to be broken on the scale $l_\phi < \xi$  in our clean samples,
while the opposite is true for the more disordered one.

The transport mechanism that leads to the observed temperature
dependence $\sigma(T)$ remains unknown. Moreover, even the nature of
the ground state of such a system is
unsettled.\cite{spivak2,spivak,spivak1,Das} Below we show that the
electron-electron interactions are extremely strong at short
distances, and decay relatively fast at larger distances due to the
screening by the metallic gate. This nature of interaction between
the delocalized carriers suggests that the holes form a
strongly-correlated liquid.

We now consider the electron-electron interaction in more details.
In HIGFETs, the metallic gate at distance $d$ from the 2D hole layer
screens the $1/r$ interaction down to $1/r^3$ when $r\gtrsim 2d$. In
our case, $d=600\,$nm for sample \#3 and $d=250\,$nm for sample \#4.
%
%

The short-distance $1/r$-interaction is indeed very strong. If
treated classically as a one-component plasma, the interaction
parameter $\Gamma = E_C/k_B T \sim 100$, corresponding to an
enormous Coulomb energy $E_C=e^2/\epsilon a\sim 10$\,K, with
$\epsilon = 13$. Since the temperature in our system is of the order
of the Fermi energy, $E_F=\hbar^2/ma^2 \sim 100\,$mK, quantum
effects may also be important. A standard estimate of the strength of
interaction is the quantum-mechanical parameter $r_s=a/a_B$, where
$a_B=\hbar^2\epsilon/me^2$ is the Bohr radius. It requires the
knowledge of the band mass $m$ that has never been measured in such
a dilute regime. Higher density cyclotron resonance measurements
give $m\simeq 0.2-0.4 \,m_e$, \cite{pan} whereas low-density
theoretical estimates (based on the Luttinger parameters)
\cite{Winkler} give $m\simeq 0.1 m_e$.
The $r_s$ value for $p=1\times 10^9\,$cm$^{-2}$
is in the range of $25 -  100$ for the mass range of $m=0.1 m_e - 0.4 m_e$.

The long-distance dipolar interaction is relatively weak for such
low densities since the $1/r^3$ potential is short-ranged in two
dimensions. Therefore, the liquid is favored over the Wigner crystal
(WC),\cite{spivak1} as the quantum fluctuations ($\sim 1/r^2$)
overcome the $1/r^3$ interaction. Meanwhile, even for a classical
system, the 2D WC melting temperature $T_m\simeq E_C/130k_{B}$ is
already low \cite{Morf}: $T_m=56\,$mK for
$p=1\times10^9\,$cm$^{-2}$. The screening further reduces\cite{Tm-screened}
$T_m$ to make the WC even harder to access. The absence
of the pinned WC in our samples is corroborated by the non-activated
transport in the linear response regime and the absence of
singularity in $\sigma(T)$.

In summary, we have observed a non-insulating behavior in the
putatively insulating regime for the hole densities down to
$7\times10^8\,$cm$^{-2}$. Our results suggest that the 2D holes form
a strongly-correlated liquid whose properties require further
investigation.

This work has benefited from valuable discussions
with I.L. Aleiner, B.L. Altshuler, P.W. Anderson,
R.N. Bhatt, and M.I. Dykman.
We also thank Stephen Chou for the use of the fabrication facilities.
The work at Princeton University is supported by US DOE grant
DEFG02-98ER45683, NSF grant DMR-0352533, and NSF MRSEC grant
DMR-0213706.

\end{document}